\RequirePackage{fix-cm}
\documentclass[twocolumn,epjc3]{svjour3}

\smartqed  
\RequirePackage{graphicx}
\journalname{Journal name}
\usepackage{cite} 
\usepackage{amsmath}
\usepackage{hyperref}
\usepackage{cite}
\usepackage{amsmath,amssymb}
\usepackage{color}

\begin{document}

\title{Scale dependent three-dimensional charged black holes in linear
and non-linear electrodynamics
}
\author{\'Angel Rinc\'on \thanksref{e1,addr1}
        \and
        Ernesto Contreras \thanksref{e2,addr2,ol} 
        \and
        Pedro Bargue\~no \thanksref{addr2}
        \and
        Benjamin Koch \thanksref{addr1}
        \and
        Grigorios Panotopoulos \thanksref{addr3}        
        \and
        Alejandro Hern\'andez-Arboleda \thanksref{addr2}
}

\thankstext{e1}{e-mail: \href{mailto:arrincon@uc.cl}{\nolinkurl{arrincon@uc.cl}} }
\thankstext{e2}{e-mail: 
\href{mailto:ej.contreras@uniandes.edu.co}{\nolinkurl{ej.contreras@uniandes.edu.co}} }
\thankstext{ol}{On leave from Universidad Central de Venezuela}
\authorrunning{Rinc\'on et al.}

\institute{Instituto de F{\'i}sica, Pontificia Universidad Cat{\'o}lica de Chile,\\ Av. Vicu{\~n}a Mackenna 4860, Santiago, Chile \label{addr1}
           \and
Departamento de F\'isica, Universidad de los Andes, \\
Apartado A\'ereo 4976, Bogot\'a, Distrito Capital, Colombia \label{addr2}
           \and
           CENTRA, Instituto Superior T{\'e}cnico, Universidade de Lisboa,\\ Av. Rovisco Pa{\'i}s 1, Lisboa, Portugal\label{addr3}        
}

\date{Received: date / Accepted: date}

\maketitle

\begin{abstract}

In the present work 
we study the scale dependence at the level of
the effective action of charged black holes in Einstein-Maxwell as well as
in Einstein-power-Maxwell theories in (2+1)-dimensional spacetimes without a
cosmological constant. We allow for scale dependence of the gravitational and 
electromagnetic couplings, and we solve the corresponding generalized field equations 
imposing the ``null energy condition". Certain properties, such as horizon structure 
and thermodynamics, are discussed in detail.


\PACS{PACS code1 \and PACS code2 \and more}

\end{abstract}

\section{Introduction}

In recent years gravity in (2+1) dimensions has attracted a lot of interest for 
several reasons. The absence of propagating degrees of freedom, its mathematical simplicity, 
the deep connection to Chern-Simons theory \cite{CS,witten1,witten2} are just a few of the reasons why to study 
three-dimensional gravity. 
In addition (2+1) dimensional black holes are a good testing ground for the four-dimensional theory, because
properties of (3+1)-dimensional black holes, such as 
horizons, Hawking radiation and black hole thermodynamics, are also present in their 
three-dimensional counterparts.  

On the other hand, the main motivation to study non-linear electrodynamics (NLED) was to overcome certain
problems of the standard Maxwell theory. In particular, non-linear electromagnetic models are introduced in order to describe
situations in which this field is strong enough to invalidate the predictions provided by the linear theory.
Originally the Born-Infeld non-linear 
electrodynamics was introduced in the 30's in order to obtain a finite self-energy of point-like 
charges \cite{BI}. During the last decades this type of action reappears in the open sector of 
superstring theories \cite{stringtheory} as it describes the dynamics of D-branes \cite{Dbranes}.
Also, these kind of electrodynamics have been coupled to gravity in order to obtain, for example, regular black holes  solutions 
\cite{ayon1998,ayon1998a, bargueno2016b}, semiclassical corrections to the black hole entropy \cite{bargueno2016a} and novel exact solutions with a
cosmological constant acting as an effective Born--Infeld cut--off \cite{BarguenoVagenas2016}. 
A particularly interesting class of 
NLED theories
is the so called power-Maxwell theory
described by a Lagrangian density of the form $\mathcal{L}(F)=F^{\beta}$, where $F=F_{\mu \nu}F^{\mu \nu}/4$ is the Maxwell invariant, and
$\beta$ is an arbitrary rational number. When $\beta=1$ one  recovers the standard linear electrodynamics, while for $\beta=D/4$ with 
$D$ being the dimensionality of spacetime, the electromagnetic energy momentum tensor is traceless \cite{cataldo2000a, chinos}. In 3 dimensions 
the generic black hole solution without imposing the traceless condition has been found in \cite{solution1}, while black hole solutions 
in linear Einstein-Maxwell theory are given in \cite{solution2,solution3}. 

Scale dependence at the level of the effective action is a generic result of quantum field theory.
Regarding quantum gravity it is well-known that its consistent formulation is still an open task. 
Although there are several approaches to quantum gravity (for an incomplete list see 
e.g. \cite{QG1,QG2,QG3,QG4,QG5,QG6,QG7,QG8,QG9} and references therein), 
most of them have something in common, namely that the basic parameters 
that enter into the action, such as the cosmological constant or Newton's constant, become scale 
dependent quantities. Therefore, the resulting effective action of most quantum gravity theories
 acquires a scale dependence. 
Those scale dependent couplings are expected to modify the properties of classical black hole backgrounds.

It is the aim of this work to study the scale dependence at the level of the effective
action of three-dimensional
charged black holes in linear (Einstein - Maxwell) and non-linear (Einstein-power-Maxwell) electrodynamics. 
We use the formalism and notation of \cite{angel,Rincon:2017ypd} where the authors applied the same technique to the BTZ 
black hole \cite{btz1,btz2}. Our work is organized as follows:
After this introduction, in the next section we present the action and the classical black hole solution
both in Einstein-Maxwell and Einstein-power-Maxwell theories. The framework and the ``null" energy
condition are introduced in sections \ref{setting} and \ref{null_energy}. The scale dependence for linear electrodynamics is presented
in the section \ref{Einstein_Maxwell}, while the corresponding solutions
for non-linear theory are given in section \ref{Einstein_power_Maxwell}. 
The discussion of our results and remarks are shown in section \ref{Discussion} whereas in section \ref{Conclusion} we summarize the main ideas and conclude. Finally, we present a brief appendix in which we show the effective Einstein field equations for an arbitrary index $\beta$ in the last section.

\section{Classical linear and non-linear electrodynamics in (2+1) dimensions}\label{clasico}

In this section we present the classical theories of linear and non-linear electrodynamics.
Those theories will then be investigated in the context of scale dependent couplings. 
The starting point is the so-called Einstein-power-Maxwell action without cosmological constant $(\Lambda_0 =0)$, 
assuming a generalized electrodynamics i.e. $\mathcal{L}(F)=C |F|^{\beta}$ which reads
\begin{align}\label{act1}
S_0 &=  \int {\mathrm {d}}^{3}x {\sqrt {-g}}\,
\bigg[\frac{1}{16 \pi G_0} R - \frac{1}{e_{0}^{2\beta}}\mathcal{L}(F) \bigg],
\end{align}
where $G_0$ is Einstein's constant, $e_0$ is the electromagnetic coupling constant, $R$ is the Ricci scalar,
$\mathcal{L}(F)$ is the electromagnetic Lagrangian density where $C$ is a constant,
$F$ is the Maxwell invariant defined in the usual way i.e. $F= (1/4)F_{\mu \nu}F^{\mu \nu}$ and
$F_{\mu \nu} = \partial_{\mu}A_{\nu} - \partial_{\nu}A_{\mu}$ 
is the electromagnetic field strength tensor.
We use the metric signature $(-, +, +)$, and natural units ($c = \hbar = k_B = 1$)
such that the action is dimensionless. Note that $\beta$ 
is an arbitrary rational number, which also appears in the exponent of the electromagnetic coupling in order to maintain the action dimensionless.
It is easy to check that the special case $\beta = 1$ reproduces the 
classical Einstein-Maxwell 
action, and thus 
the standard electrodynamics is recovered. For $\beta \neq 1$ one can obtain Maxwell-like solutions.
In the following we shall consider both cases: first when $\beta = 3/4$, since it 
is this value that allows us to obtain a trace-free electrodynamic tensor, precisely as in the four-dimensional standard Maxwell theory, and second when $\beta=1$ because is the usual electrodynamics in 2+1 dimensions. 
In both cases one obtains the same classical equations of motion, which are given 
by Einstein's field equations
\begin{align}\label{Gmunu}
G_{\mu \nu} &= \frac{8 \pi G_0}{e_0^{2\beta}} T_{\mu \nu }.
\end{align}
The energy momentum tensor $T_{\mu \nu}$ is associated to the electromagnetic field strength $F_{\mu \nu}$ through
\begin{align}\label{TNL}
T_{\mu \nu} &= \mathcal{L}_{F} g_{\mu \nu} - \mathcal{L}(F) F_{\mu \gamma} F_{\nu}\ ^{\gamma},
\end{align}
remembering that $\mathcal{L}_F=d\mathcal{L}/dF$. In addition, for static 
circularly symmetric solutions the electric 
field $E(r)$ is given by
\begin{align}\label{Fmunu}
F_{\mu \nu} &= (\delta_{\mu}^{r} \delta_{\nu}^{t} - \delta_{\nu}^{r} \delta_{\mu}^{t})E(r).
\end{align}
For the metric circular symmetry implies
\begin{equation}\label{metric}
ds^2 = -f(r) dt^2 + g(r) dr^2 + r^2 d \phi^2.
\end{equation}
Note that in the classical case one finds that $g(r)=f(r)^{-1}$.
Finally, the equation of motion for the Maxwell field  $A_{\mu}(x)$ reads
\begin{align} \label{decov}
D_{\mu}\left(\frac{\mathcal{L}_{F}F^{\mu\nu}}{e_0^{2\beta}}\right)& = 0.
\end{align}
With the above in mind, for charged black holes one only needs to
 determine the set of functions $\{f(r), E(r)\}$. Using Einstein's field 
equations \ref{Gmunu} and Eq. \ref{decov} combined with Eq \ref{Fmunu} and the definition of $\mathcal{L}_F$, one obtains 
the classical electric field as well as the lapse function $f(r)$.

\noindent It is possible to determine the electric field as well as the lapse function without assuming a particular value 
for $\beta$ for classical solutions, however we will focus on two of them. First, the Einstein-Maxwell case is in itself 
interesting due 
to its relation with the four-dimensional case. On the other hand, the Einstein-power-Maxwell case with $\beta = 3/4$ is a desirable one due to 
a remarkable property: it has a null trace, which is also present in the four-dimensional case. The general treatment for any value of $\beta$ can be found in the appendix \ref{Appendix}.
\\
 
\subsection{Einstein-Maxwell case}
The classical (2+1)-dimensional Einstein-Maxwell black hole solution  ($\beta=1$) is given by
\begin{align}
f_0(r) &=-M_0 G_0-\frac{1}{2}\frac{Q_0^2}{e_{0}^{2}} \ln\bigg(\frac{r}{\tilde{r}_{0}}\bigg),
\\
E_0(r) &= \frac{Q_0}{r} e_{0}^{2},
\end{align} 
where $M_0$ is the mass and $Q_0$ is the electric charge of the black hole
and $\tilde{r}_{0}$ stands for the radius where the electrostatic potential 
vanishes. The apparent horizon $r_0$ is obtained by demanding that $f_0(r_0)=0$, which reads 
\begin{align}
r_0 &= \tilde{r}_{0} e^{-\frac{2 M_0 G_0 e^{2}_{0}}{Q_0^2}},
\end{align}
and rewriting the lapse function using the apparent horizon one gets
\begin{align}
f_0(r) &= - \frac{Q_0^2}{2 e_{0}^{2}} \ln\left(\frac{r}{r_0}\right).
\end{align}
Black holes have thermodynamic behaviour.
Here, the 
Hawking temperature $T_0$, the Bekenstein-Hawking entropy $S_0$, as well as the heat capacity $C_0$ are found to be
\begin{align}
T_0(r_0) &= \frac{1}{4 \pi} \Bigg|  \frac{Q_0^2}{2 e_{0}^{2}r_0}\Bigg|,
\\
S_0(r_0) &= \frac{\mathcal{A}_H(r_0)}{4G_0},
\\
C_0(r_0) &= T\frac{\partial S}{\partial T}\Bigg|_{Q} = - S_0(r_0).
\end{align}
Note that $\mathcal{A}_H(r_0)$ is the horizon area which is given by
\begin{align}
\mathcal{A}_{H}(r_0) &= \oint  {\mathrm {d}}x \sqrt{h} = 2 \pi r_0, 
\end{align}

\subsection{Einstein-power-Maxwell case}\label{pem}
Solving Einstein's field equations for $\beta=3/4$, 
the lapse function $f(r)$ and the electric field $E(r)$ are found to be
\begin{align}
f_0(r) &=\frac{4 G_0 Q_0^2}{3 r} - G_0 M_0 \label{fdr},
\\
E_0(r) &= \frac{Q_0}{r^2}. \label{edr}
\end{align}
It is worth mentioning that, unlike in the previous section, the solutions here considered do not contain the electromagnetic coupling. 
This is due to the fact that a dimensional analysis on the action (\ref{act1}) for $\beta=3/4$ 
reveals that the electric charge is dimensionless in this case. As a consequence, we can set the electromagnetic coupling to unity without 
affecting the classical action.
\\
At classical level a horizon is present, and it is computed by requiring that $f(r_0)=0$, which reads
\begin{align}\label{rclasico34}
r_0 &= \frac{4}{3}\frac{Q_0^2}{M_0}.
\end{align}
Expressing the mass $M_0$ in terms of the horizon one obtains
\begin{equation}
f_0(r) = \frac{4}{3}G_0 Q_0^2 \Bigg[ \frac{1}{r} - \frac{1}{r_0} \Bigg].
\end{equation}
Classical thermodynamics plays a crucial role since it provides us with valuable information about the underlying black hole physics. 
The Hawking temperature $T_0$, the 
Bekenstein-Hawking entropy $S_0$ as well as the heat capacity $C_0$ are given by
\begin{align}
T_0(r_0) &= \frac{1}{4\pi}
\Bigg| \frac{M_0 G_0}{r_0} \Bigg|, \label{Tclassic}
\\
S_0(r_0) &= \frac{\mathcal{A}_H(r_0)}{4 G_0},\label{Sclassic}
\\
C_{0}(r_0) &= - \frac{\mathcal{A}_H(r_0)}{4 G_0}. \label{CQclassic}
\end{align}
In agreement with the notation in the previous section,  $\mathcal{A}_{H}(r_0)$ is the so-called horizon area.

\section{Scale dependent couplings and scale setting}\label{setting}

This section summarizes the equations of motion for the scale dependent Einstein-Maxwell and Einstein-power-Maxwell theories. 
The notation follows closely \cite{Koch:2014joa} as well as \cite{angel,Rincon:2017ypd}. 

\noindent The scale dependent couplings of the theories are i) the gravitational coupling $G_k$, and ii) the electromagnetic 
coupling $1/e_k$. Furthermore, there are three independent fields, which are the metric $g_{\mu \nu}(x)$, the electromagnetic
four-potential $A_{\mu}(x)$, and the scale field $k(x)$. 
The effective action for the non-linear electrodynamics reads
\begin{align}\label{Effective_Actiom}
\Gamma[g_{\mu \nu}, k] &=  \int {\mathrm {d}}^{3}x {\sqrt {-g}}\,
\bigg[\frac{1}{2\kappa_k} R - \frac{1}{e_{k}^{2\beta}}\mathcal{L}(F) \bigg],
\end{align}
The equations of motion for the metric $g_{\mu \nu}(x)$ are given by
\begin{align}
G_{\mu\nu} &= \frac{\kappa_k}{e^{2\beta}_k}T^{\text{effec}}_{\mu\nu},
\end{align}
with
\begin{align}
T^{\text{effec}}_{\mu\nu} &= T^{\text{EM}}_{\mu\nu} - \frac{e^{2\beta}_k}{\kappa_k} \Delta t_{\mu \nu}.
\end{align}
Note that $T^{\text{EM}}_{\mu\nu}$ is given by Eq. \ref{TNL}, $\kappa_k=8 \pi G_k$ is the Einstein constant and the additional object $\Delta t_{\mu \nu}$ is defined as follows
\begin{align}
\Delta t_{\mu\nu} &= G_k \Bigl(g_{\mu \nu} \square - \nabla_{\mu} \nabla_{\nu}
\Bigl)G_k^{-1}.
\end{align}
The equations of motion for
the four-potential $A_{\mu}(x)$ taking into account the running of $e_k$ are
\begin{align} \label{decovcoupling}
D_{\mu}\left(\frac{\mathcal{L}_{F}F^{\mu\nu}}{e_k^{2\beta}}\right)& = 0.
\end{align}
It is important to note that
since the renormalization scale $k$ is actually not constant 
any more,
this set of equations of motion do not close consistently by itself.
This implies that the stress energy tensor is most likely not conserved
for almost any choice of functional dependence $k=k(r)$.
This type of scenario has largely been explored in the context
of renormalization group improvement of black holes in asymptotic safety scenarios \cite{Bonanno:2000ep,Bonanno:2006eu,Reuter:2006rg,Reuter:2010xb,Falls:2012nd,Cai:2010zh,Becker:2012js,Becker:2012jx,Koch:2013owa,Koch:2013rwa,Ward:2006vw,Burschil:2009va,Falls:2010he,Koch:2014cqa,Bonanno:2016dyv}.
The loss of a conservation laws comes from the fact that there is one consistency equation missing.
This missing equation can be obtained from varying the effective action (\ref{Effective_Actiom}) with respect to the scale field $k(r)$, i.e.
\begin{equation}\label{vary}
\frac{d}{d k} \Gamma[g_{\mu \nu},k] =0,
\end{equation}
which can thus be understood
as variational scale setting procedure \cite{Reuter:2003ca,Koch:2010nn,Domazet:2012tw,Koch:2014joa,Contreras:2016mdt}.
The combination of (\ref{vary}) with the above equations of motion guarantees the
conservation of the stress energy tensors.
A detailed analysis of the split symmetry within the functional renormalization group equations, 
supports this approach of dynamic scale setting \cite{Percacci:2016arh}.

The variational procedure (\ref{vary}), however, 
requires the knowledge of the exact beta functions of the problem.
Since in many cases the precise form of the beta functions 
is unknown (or at least unsure) one can, for
the case of simple black holes, impose a null energy condition
and solve for the couplings $G(r),\, \Lambda(r),\, e(r)$ directly 
\cite{Contreras:2013hua,Koch:2015nva,angel,Rincon:2017ypd}.
This philosophy of assuring the consistency of the equations by imposing
a null energy condition will also be applied in the following study on
Einstein-Maxwell and Einstein-power-Maxwell black holes.

\section{The null energy condition}\label{null_energy}

The so-called Null Energy Condition (hereafter NEC) is the less restrictive of the usual energy conditions (dominant, weak, strong, and null), and it 
helps us to obtain desirable solutions of Einstein's field 
equations \cite{Rubakov:2014jja,Wald:1984rg}. 
Considering a null vector $\ell^{\mu}$, the NEC is applied on the matter stress energy tensor such as
\begin{align}
T^{m}_{\mu \nu} \ell^{\mu} \ell^{\nu} \geq 0.
\end{align}
The application of such a condition was appropriately implemented in Ref. \cite{angel} inspired by the Jacobson idea \cite{Jacobson:2007tj}. Note 
that in proving fundamental black hole theorems, such as the no hair theorem \cite{Heusler:1996ft} and the 
second law of black hole thermodynamics \cite{Bardeen:1973gs}, the NEC is, indeed, required.
For scale dependent couplings, one requires that the aforementioned condition is not violated and, therefore, the 
NEC is applied on the effective stress energy tensor for a special null vector $\ell^{\mu}=\{f^{-1/2}, f^{1/2}, 0\}$ such as
\begin{align}
T^{\text{effec}}_{\mu \nu} \ell^{\mu} \ell^{\nu} = \bigg(T^{\text{EM}}_{\mu \nu} - \frac{e^{2\beta}_k}{\kappa_k} \Delta t_{\mu \nu}\bigg) \ell^{\mu} \ell^{\nu} \geq 0.
\end{align}
In addition, the left hand side (LHS) is null as well as $T^{\text{EM}}_{\mu \nu}\ell^{\mu} \ell^{\nu} =0$ and the condition 
reads
\begin{align}\label{Delta_t}
\Delta t_{\mu \nu} \ell^{\mu} \ell^{\nu} = 0.
\end{align}
One should note that Eq. \ref{Delta_t} allows us to obtain the gravitational coupling $G(r)$ easily by solving the differential equation 
\begin{align}
G(r)\frac{d^2 G(r)}{dr^2} - 2 \left(\frac{d G(r)}{dr}\right)^2=0,
\end{align}
which leads to 
\begin{align}
G(r) = \frac{G_0}{1 + \epsilon r}.
\end{align}
The NEC allows us to decrease the number of degrees of freedom, and thus it becomes an important tool for scale dependent black hole problems.

\section{Scale dependence in Einstein-Maxwell theory} \label{Einstein_Maxwell}

In order to get insight into non-linear electrodynamics
regarding the running of couplings, one first has to discuss the effects of scale dependence in linear electrodynamics. With this in mind, one also needs to determine the set of four functions $\{G(r), E(r), f(r), e(r)^2\}$ which 
are obtained by combining Einstein's effective equations of motion with the NEC taking into account the EOM for the four-potential $A_{\mu}$.

\subsection{Solution}
\begin{figure}[ht!]
\centering
\includegraphics[width=\linewidth]{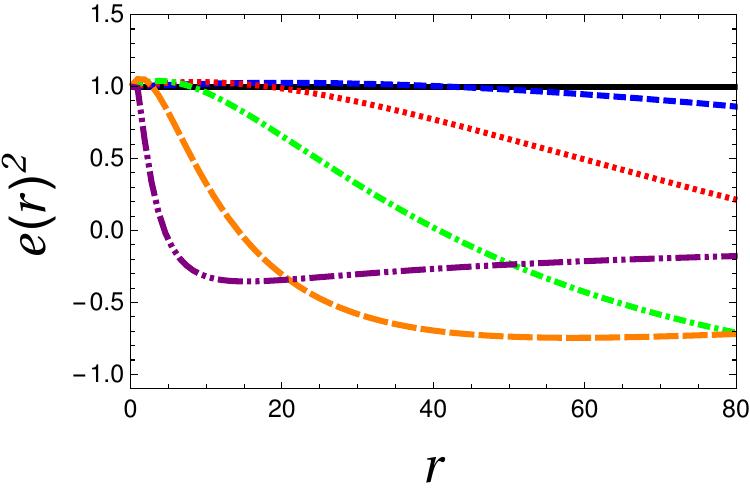}
\caption{\label{er2_0} Electromagnetic coupling $e(r)^2$ for $\epsilon=0$ (black solid line), $\epsilon=0.0025$ (short dashed blue line), $\epsilon=0.007$ 
(dotted red line), $\epsilon=0.02$ (dotted dashed green line), $\epsilon=0.08$ (long dashed orange line) and  $\epsilon=0.5$ (double dotted dashed purple line). The other values have been taken as unity.
}
\end{figure}

The solution for this scale dependent black hole is given by
\begin{align} \label{setsoluII}
G(r) &= \frac{G_0}{1 + \epsilon r}, \nonumber
\\
E(r) &= \frac{Q_0}{r}e(r)^2,\nonumber \\
f(r) &= -\frac{G_0 M_0}{(r \epsilon +1)^2}
-
\frac{Q_0^2}{2 e_{0}^{2}}
\frac{ (\ln (r/\tilde{r}_{0}) +  r \epsilon)}{ (r \epsilon +1)^2},
\\
e(r)^2 &= e_0^2 \Bigg[
\frac{1}{(1 + r\epsilon)^3} 
+ 
4 \ \frac{ r \epsilon}{(1 + r\epsilon)^3} \nonumber
\\
& \ \  - 
\Bigg(4 M_0 G_0 - 5 Q_0^2 + 2 Q_0^2 \ln\bigg(\frac{r}{\tilde{r}_{0}}\bigg)\Bigg)\frac{r^2\epsilon^2}{(1 + r\epsilon)^3}
\Bigg] .\nonumber
\end{align}
where the integration constants are chosen such as the classical Einstein-Maxwell (2+1)-dimensional black hole is recovered 
according to \cite{Martinez:1999qi}. 
It is relevant to say that the gravitational coupling $G(r)$ is obtained by taking advantage of NEC, while the electric field
$E(r)$ is given by the covariant derivative \ref{decovcoupling}, which depends on the electromagnetic coupling constant $e(r)$. Besides, the lapse 
function $f(r)$ and the coupling $e(r)$ are directly obtained by using Einstein's effective field equations combined with the solutions for $E(r)$ and $G(r)$.
In addition, our solution reproduces the results of the classical theory in the limit $\epsilon \rightarrow 0$, i.e.
\begin{align}
\lim_{\epsilon \rightarrow 0} G(r) &= G_0, \nonumber
\\
\lim_{\epsilon \rightarrow 0} E(r) &= \frac{Q_0}{r} e_{0}^{2},\nonumber \\
\lim_{\epsilon \rightarrow 0} f(r) &= - G_0 M_0
-
\frac{Q_0^2}{2 e_{0}^{2}}\ln\bigg(\frac{r}{\tilde{r}_{0}}\bigg),
\\
\lim_{\epsilon \rightarrow 0} e(r)^2 &= e_0^2.\nonumber
\end{align}
which justifies the naming of the constants aforementioned $\{G_0, M_0, Q_0, e_0\}$ in terms of their meaning in the absence of scale dependence
\cite{angel}, as it should be.
Besides, the parameter $\epsilon$ controls the strength of the new scale dependence effects, and therefore it is useful to
treat it as a small expansion parameter as follows
\begin{align}
G(r) &\approx G_0 \Bigl[1 -r \epsilon + \mathcal{O}(\epsilon^2)\Bigl],
\\
E(r) &\approx \frac{Q_0}{r}e_{0}^{2}\bigg[1 + \epsilon r + \mathcal{O}(\epsilon^2)\bigg],
\\
f(r) &\approx f_0(r) +
\bigg[
2 G_0 M_0 - \frac{1}{2}\frac{Q_0^2}{e_{0}^{2}} 
\\
& \hspace{1.3cm}+ \frac{Q_0^2 }{e_{0}^{2}} \ln\bigg(\frac{r}{\tilde{r}_{0}}\bigg)
 \bigg]r\epsilon 
 + \mathcal{O}(\epsilon^2),\nonumber
\\
e(r)^2 &\approx e_0^2 \bigg[1 + \epsilon r + \mathcal{O}(\epsilon^2)\bigg].
\end{align}
\begin{figure}[ht!]
\centering
\includegraphics[width=\linewidth]{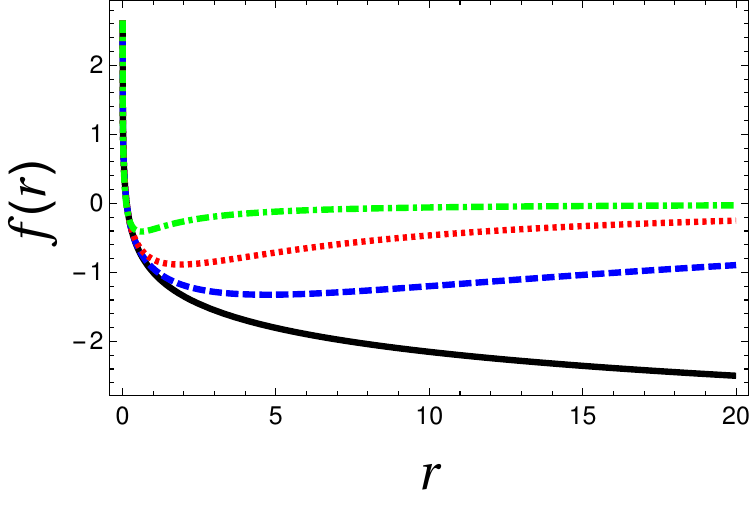}
\caption{\label{fr_0} Lapse function $f(r)$ for $\epsilon=0$ (black solid line), $\epsilon=0.04$ (blue dashed line), $\epsilon=0.15$ 
(dotted red line) and $\epsilon=1$ (dotted dashed green line). The values for the rest of the parameters have been taken as unity.
}
\end{figure}
In figure \ref{fr_0} the lapse function $f(r)$ is shown for different values of $\epsilon$ in comparison to the classical (2+1)-dimensional 
Einstein-Maxwell solution. The figure shows that the scale dependent solution for small $\epsilon\cdot r$ values is consistent with the classical 
case. However, when $\epsilon\cdot r$ becomes sufficiently large, 
a deviation from the classical solution appears. 
The electromagnetic coupling $e(r)$ is shown in Figure \ref{er2_0} for different values of $\epsilon$. Note that when
$\epsilon$ is small the classical case is recovered, but when $\epsilon$ increases the electromagnetic coupling tends to decrease until 
it is stabilized.

\subsection{Asymptotic behaviour}
In this subsection a few invariants need to be revisited. In particular we will focus on the Ricci 
scalar $R$ and the 
Kretschmann scalar $\mathcal{K}$. Both of them are relevant in order to check if some additional divergences appear. 
For the static and circularly symmetric metric we have considered, 
the Ricci scalar is given by
\begin{align}
R & = -f''(r)-\frac{2 f'(r)}{r},
\end{align}
or more precisely
\begin{align}\label{Rmaxwell}
R &= \frac{Q_0^2}{2r^2 e_{0}^{2}(1+r\epsilon)^4} 
- \frac{8 M_0 G_0 e_{0}^{2} + 4 Q_0^2\ln(r/\tilde{r}_{0})}{2r^2 (1+r\epsilon)^4  e_{0}^{2}} r \epsilon
\\
&+ \frac{4 M_0 G_0 e_{0}^{2} -7 Q_0^2 
+ 2 Q_0^2 \ln(r/\tilde{r}_{0})}{2r^2 (1+r\epsilon)^4 e^{2}_{0}} (r \epsilon)^2. \nonumber
\end{align}
We require that classically the Ricci scalar reads
\begin{align}\label{Rmaxwellclassic}
R_0 &= \frac{1}{2}\frac{Q_0^2}{e_{0}^{2}r^2}.
\end{align}
Considering $r$ values close to zero one obtains
\begin{align}
R \approx R_0 \Bigg[
1 - 
\Bigg[\frac{8M_0G_0 e_{0}^{2}}{Q_0^2} + 4\ln\bigg(e\frac{r}{\tilde{r}_{0}}\bigg)\Bigg]r\epsilon 
+ \cdots
\Bigg].
\end{align}
Thus, upon comparing Eq.\ref{Rmaxwell} with Eq.\ref{Rmaxwellclassic} we observe 
that the scale dependent effect strongly distorts this invariant. Despite that, for small values of $r$ the standard case $R_0$ 
is recovered. In the same way, one expects that $\epsilon$ 
should be small, therefore one can expand the Ricci scalar around $\epsilon=0$ but the solution is exactly the same reported for $r \ll 1$.
Regarding the Kretschmann scalar, it is computed to be
\begin{align}
\mathcal{K} &= R_{\mu \nu \alpha \beta}R^{\mu \nu \alpha \beta}.
\end{align}
Thus, when $\epsilon$ is small the Kretschmann scalar reads
\begin{align}
\mathcal{K} \approx \mathcal{K}_0 \Bigg[
1
-
\Bigg[\frac{16M_0G_0 e_{0}^{2}}{3Q_0^2} + 
\frac{8}{3}\ln\bigg(\frac{r}{\tilde{r}_{0}}\bigg)
\Bigg]r\epsilon
\Bigg] + \cdots
\end{align}
Note that the classical result for this invariant is indeed $\mathcal{K}_0=3Q_0^4/4r^4$, which coincides with our solution when 
$\epsilon \rightarrow 0$.

\noindent The other regime of asymptotic behaviour can be studied in a large radius expansion $r\to\infty$. In this limit the lapse function $f(r)$
decays as $r^{-1}$ which disagrees with the classical result shown in Eq.\ref{fdr}.
On the other hand, the electromagnetic coupling $e(r)$ also tends to zero as $r^{-1}$
in contrast with the expected result, $e_{0}$.
Finally, one obtains that $E(r)\sim r^{-2}$, $R\sim r^{-4}$ and $\mathcal{K}\sim r^{-6}$, all of them going to zero as expected. However,
it can be shown that these functions decay faster than those corresponding to the classical solutions. 
In fact, in absence of running coupling, a straightforward calculation reveals that $E(r)\sim r^{-1}$, $R\sim r^{-2}$ and $\mathcal{K}\sim r^{-4}$.

\subsection{Horizons}
\begin{figure}[ht!]
\centering
\includegraphics[width=\linewidth]{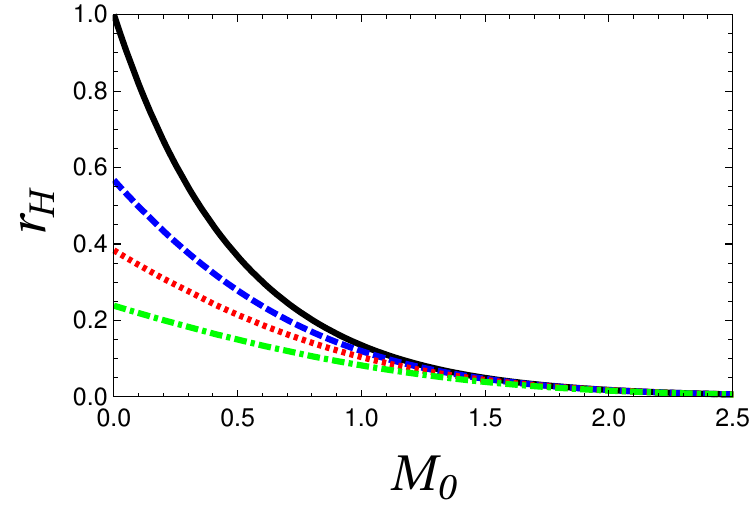}
\caption{\label{rH_0} 
Black hole horizons $r_H$ as a function of the mass $M_0$ for 
 $\epsilon=0$ (black solid line), $\epsilon=1$ (blue dashed line), $\epsilon=2.5$ (dotted red line) and $\epsilon=6$ (dotted dashed green line). The values of the rest of the parameters have been taken as unity.
}
\end{figure}

The apparent horizon occurs when the lapse function vanishes, i.e. $f(r_H)=0$. 
Thus, this Einstein-Maxwell black hole solution represents a non trivial deviation from the classical solution which is manifest when we compare our solution with the corresponding black hole solution 
without the scale dependence. Here, the horizon read
\begin{align}\label{rHmaxwell}
r_H &= \frac{1}{\epsilon }W \left(\epsilon  e^{-\frac{2 G_0 M_0 e_{0}^{2}}{Q_0^2}}\right),
\end{align}
where $W(\cdot)$ is the so-called Lambert-$W$ function, which is a set of functions, namely the branches of the inverse relation of the function $Y(r\epsilon) = r\epsilon e^{r\epsilon}$ with $r\epsilon$ being a complex number. In particular, Eq \ref{rHmaxwell} is also the principal solution for $r\epsilon$. In Figure \ref{rH_0} the scale dependent effect on horizon is shown. We can see that the deviation from the classical case is also evident for small $M_0$ values.

\noindent In addition, one can expand the horizon around $\epsilon = 0$ obtaining the classical solution plus corrections i.e.
\begin{align}
r_H &\approx r_0\Bigg[1 - \epsilon r_0 + \mathcal{O}(\epsilon^2)\Bigg].
\end{align}

\subsection{Thermodynamic properties}
\begin{figure}[ht!]
\centering
\includegraphics[width=\linewidth]{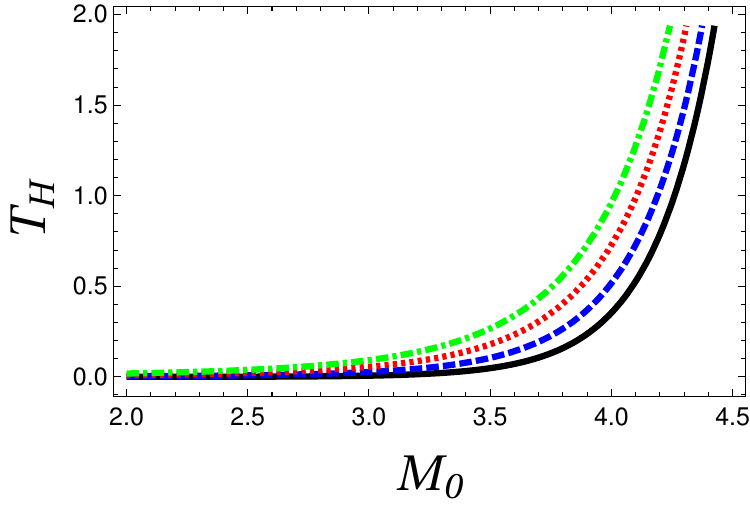}
\caption{\label{TH_0} 
The Hawking temperature $T_H$ as function of the classical mass $M_0$ for $\epsilon=0$ (black solid line), $\epsilon=750$ (blue dashed line), $\epsilon=1800$ (dotted red line) and $\epsilon=3000$ (dotted dashed green line). The other values of the rest of 
the parameters
have been taken as unity. 
Note that the vertical axis is scaled $1:10^{6}$
}
\end{figure}

After having gained experience on the horizon structure one can now move towards the usual thermodynamic properties associated with our solution shown at Eq. \ref{setsoluII}. Thus, the Hawking temperature of the black hole assuming the ansatz \ref{metric} is given by
\begin{align}
T_H(r_H) &= \frac{1}{4 \pi} \Bigg|\lim_{r\rightarrow r_H} \frac{\partial_r g_{tt}}{\sqrt{-g_{tt}g_{rr}}} \Bigg|,
\end{align}
i.e. 
\begin{align}
T_H(r_H) &= \frac{1}{4 \pi}\Bigg|\frac{Q_0^2}{2 r_H(1 + \epsilon r_H) e_{0}^{2}}\Bigg|.
\end{align}
Taking advantage of the fact that the integration constant $\epsilon$ should be small, one can expand around $\epsilon = 0$ to get the well-known
Hawking temperature (at leader order) i.e.
\begin{align}
T_H(r_H) &\approx T_0(r_0) \Bigl|1 + \epsilon r_0  + \mathcal{O}(\epsilon^2)\Bigl|.
\end{align}
In Figure \ref{TH_0} we show the effective temperature which takes into account the running coupling effect.

\begin{figure}[ht!]
\centering
\includegraphics[width=\linewidth]{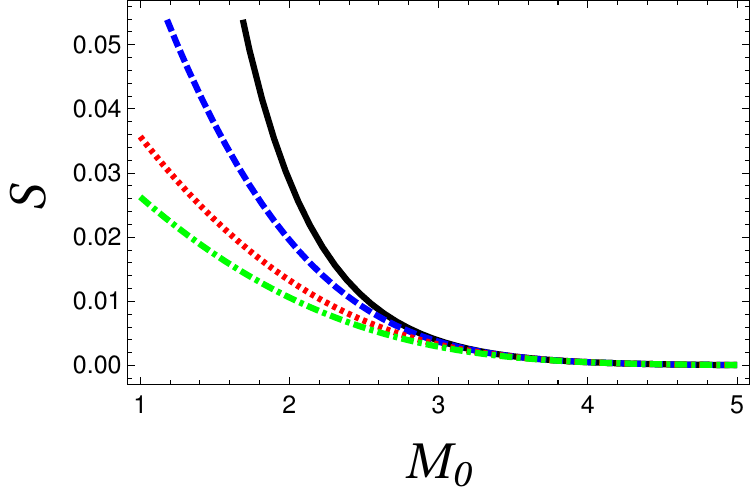}
\caption{\label{S_0} 
The Bekenstein-Hawking entropy $S$ as function of classical mass $M_0$ for $\epsilon=0$ (black solid line), $\epsilon=200$ (blue dashed line), $\epsilon=600$ (dotted red line) and $\epsilon=1000$ (dotted dashed green line). The other values have been taken as unity.
}
\end{figure}

\noindent 
Moreover, the Bekenstein-Hawking entropy for his black hole is
\begin{align}
S &=\frac{\mathcal{A}_H(r_H)}{4G(r_H)} = S_0(r_H)(1 + \epsilon r_H),
\end{align}
and assuming small values of $\epsilon$ one can expand to get
\begin{align}
S &\approx S_0(r_0) \Bigg[1 -\frac{1}{2}(\epsilon r_0)^2 +\mathcal{O}(\epsilon^3) \Bigg].
\end{align}
In Figure \ref{S_0} below we show the entropy for our (2+1)-dimensional Einstein-Maxwell scale dependent black hole. It is clear that the running effect is dominant when $\epsilon$ is not small, while for large values of $M_0$ the effect is practically zero.

\noindent Finally, the heat capacity is computed in the usual way i.e.:
\begin{align}
C_Q &= T \ \frac{\partial S}{\partial T}\Bigg|_{Q} ,
\end{align}
which read
\begin{align}
C_Q &= -S_0(r_H)(1 + \epsilon r_H ).
\end{align}
The classical case is, of course, recovered in the $\epsilon \rightarrow 0$ limit.
Due to a weak $\epsilon$ dependence it was necessary to 
plot the figure with very large values of $\epsilon$ in order to generate a visible effect. 
The scale dependent effect is notoriously small for those quantities.

\subsection{Total charge}
The electric field is parametrized through the total char\-ge $Q$, but in our previous discussion $Q_0$ only denotes
an integration constant which coincides with the charge of the classical theory. In general, we need to compute the total charge by the following relation \cite{Carroll:2004st}
\begin{align}
Q &= \int \sqrt{-g} \mathrm{d} \Omega \Bigg(\frac{\mathcal{L}_F F_{\mu \nu}}{e_k^{2 \beta}}\Bigg)n^{\mu} \sigma^{\nu},
\end{align}
where $n^{\mu}$ and $\sigma^{\nu}$ are the unit spacelike and timelike vectors normal to the hypersurface of radius $r$, and they are given by $n^{\mu}=(f^{-1/2},0,0)$ and $\sigma^{\nu}=(0,f^{1/2},0)$ as well as $\sqrt{-g}\mathrm{d} \Omega = r\mathrm{d} \phi$. Making use of these we obtain
\begin{align}
Q &= 2\pi Q_{0},
\end{align}
which is proportional to the classical value and has no $\epsilon$ dependence.

\section{Einstein-power-Maxwell scale dependence} \label{Einstein_power_Maxwell}
This section is devoted to the study of a $(2+1)$ scale dependent gravity coupled to a power-Maxwell source. 
As mentioned before, the case $\beta=3/4$ leads to a dimensionless electromagnetic 
coupling which was set to the unity in the section \ref{pem}. However, if one considers a scale dependent gravity, the 
electromagnetic coupling has a non-trivial scale dependence. Therefore, in this
section we shall hold the electromagnetic coupling dependence of the action \ref{act1}. In this way,
the solution consists of a set of four functions
$\{G(r), E(r), f(r), e(r)^3\}$, which are obtained 
by combining Einstein's effective equations of 
motion with the NEC taking advantage of the EOM for the four-potential $A_{\mu}$. In what follows
we shall obtain the solutions of the system in terms of the functions mentioned above.

\subsection{Solution}
The integration constants have been chosen such as the scale dependent solution reduces to the classical NLED case when the appropriate limit
is taken. Thus, our solution reads
\begin{align} \label{setsolu}
G(r) &= \frac{G_0}{1 + \epsilon r}, \nonumber
\\
E(r) &= \frac{Q_0}{r^2}\left(\frac{e(r)}{e_0}\right)^3,\\
f(r) &= \frac{4 G_0 Q_0^2}{3 r (r \epsilon +1)^3}
-
\frac{M_0 G_0 \left(r^3 \epsilon ^2+3 r^2 \epsilon +3 r\right)}{3 r (r \epsilon +1)^3},\nonumber
\\
e(r)^3 &= e_0^3 \Bigg[\frac{(2 r \epsilon  (3 r \epsilon +2)+1)}{(r \epsilon +1)^4} 
-
\frac{M_0 r^3 \epsilon ^2 (r \epsilon +4)}{4 Q_0^2 (r \epsilon +1)^4}
\Bigg].\nonumber
\end{align}
In the limit $\epsilon \rightarrow 0$ we obtain
\begin{align}
\lim_{\epsilon \rightarrow 0} G(r) &= G_0, \nonumber
\\
\lim_{\epsilon \rightarrow 0} E(r) &= \frac{Q_0}{r^2},\\
\lim_{\epsilon \rightarrow 0} f(r) &= \frac{4 G_0 Q_0^2}{3 r}- G_0 M_0,\nonumber
\\
\lim_{\epsilon \rightarrow 0}  e(r)^3 &= e_0^3 .\nonumber
\end{align}
Note that if we set $e_{0}=1$, the classical solution in section \ref{pem} is recovered. Even more, if one demands that $G_0 = 1$ (which is the standard lore) then we are in complete agreement with the classical solution given at Ref. \cite{Cataldo:2000we}.

\subsection{Asymptotic behaviour}
The asymptotic behaviour of this solution can be studied by computing geometrical invariants i.e. the Ricci scalar, which for
our solution is
\begin{align}
R &= -4 G_0 \epsilon \Bigg[\frac{ M_0 + 4 Q^2 \epsilon}{r (r \epsilon +1)^5}\Bigg],
\end{align}
where the classical case (with a null cosmological constant) is clearly $R_0=0$. For $r \rightarrow 0$ one obtains
\begin{align}
R &\approx -4 G_0 \epsilon \Bigg[ \frac{M_0 + 4 Q_0^2 \epsilon}{r} \Bigg] + \mathcal{O}(r).
\end{align}
We observe that the Ricci scalar is altered in presence of scale dependent coupling. In addition, one note 
that an unexpected $r^6$ divergence appears, which is controlled by $\epsilon$.
Another geometrical invariant is the Kretschmann scalar $\mathcal{K}$ which is given by
\begin{align}
\mathcal{K} &= R_{\mu \nu \alpha \beta}R^{\mu \nu \alpha \beta}.
\end{align}
For $r \rightarrow 0$ one can obtain the first terms which are
\begin{align}
\mathcal{K} &\approx \frac{32 G_0^2 Q_0^4}{3 r^6} 
\Bigg[ 1
-
\Bigg(
\frac{M_0}{Q_0^2} \epsilon + 4 \epsilon^2
\Bigg) r^2
\Bigg]
 + \mathcal{O}(r^{-3}).
\end{align}
Taking into account that the $\epsilon$ should be small we have
\begin{align}
\mathcal{K} &\approx \frac{32 G_0^2 Q_0^4}{3 r^6} 
\Bigg[
1-\frac{M_0 r^2}{Q_0^2} \epsilon
 + \mathcal{O}(\epsilon^2)
\Bigg],
\end{align}
where the standard value $\mathcal{K}_0$ has been obtained demanding that $\epsilon$ goes to zero. Classically, the Ricci scalar 
for null cosmological constant is identically zero, however in presence of scale dependent couplings it exhibits a singularity. The Kretschmann 
scalar exhibits a singularity at $r \rightarrow 0$ for both the classical and the scale dependent case. 
On the other hand, the opposite regime of asymptotic behaviour is studied in the large radius expansion $r \rightarrow \infty$ both for the Ricci and the Kretschmann scalar. The Ricci scalar 
as well as the Kretschmann scalar are asymptotically close to zero.

\begin{figure}[ht!]
\centering
\includegraphics[width=\linewidth]{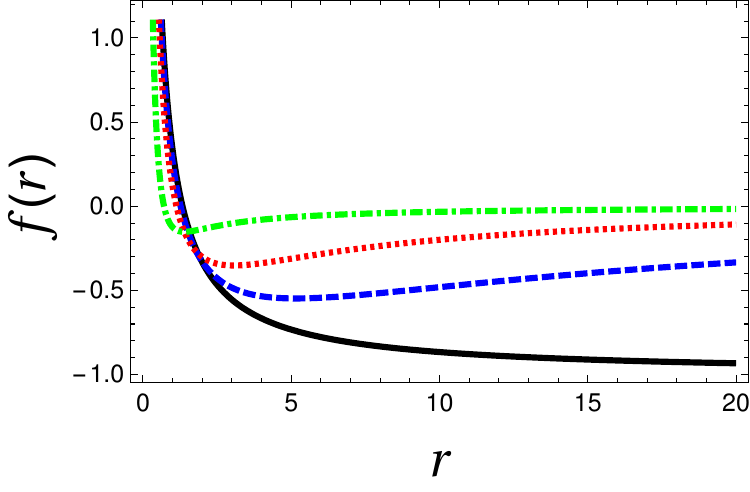}
\caption{\label{fr_I} Lapse function $f(r)$ for $\epsilon=0$ (black solid line), $\epsilon=0.04$ (blue dashed line), $\epsilon=0.15$ (dotted red line) and $\epsilon=1$ (dotted dashed green line). The values of the rest of the parameters have been taken as unity.
}
\end{figure}

\begin{figure}[ht!]
\centering
\includegraphics[width=\linewidth]{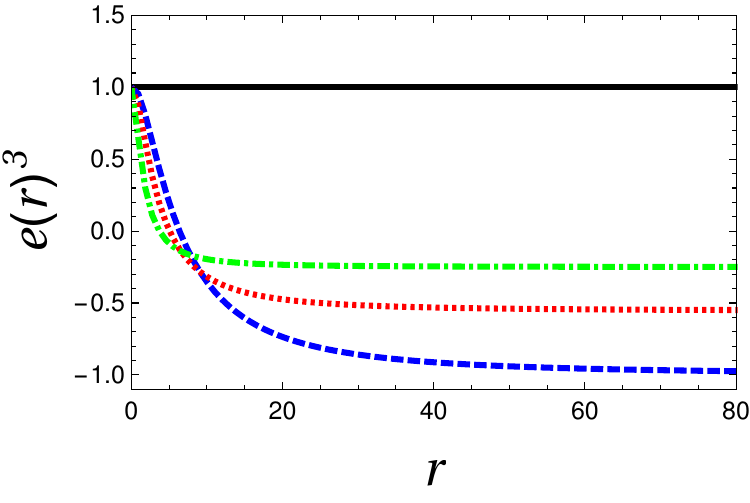}
\caption{\label{er3_I} Electromagnetic coupling $e(r)^3$ for $\epsilon=0$ (black solid line), $\epsilon=0.25$ 
(dashed blue line), $\epsilon=0.45$ (dotted red line) and $\epsilon=1$ (dotted dashed green line). The values of the rest of the parameters have been taken as unity.
}
\end{figure}

\noindent Regarding the limit $r \rightarrow \infty$ the lapse function 
goes as $r^{-1}$ in agreement with the  asymptotic behaviour of the classical solution.
In addition, note the unusual behaviour of the electromagnetic coupling in the light of scale dependent framework in Fig. \ref{er3_I}. Starting from $e_0^3$ the electromagnetic coupling decays softly and it stabilizes when 
\begin{align}
\lim_{r\rightarrow \infty} e(r)^3 &= - \bigg(\frac{1}{3 r_0 \epsilon}\bigg) e_0^3,
\end{align}
instead of reach the classical value.
The electric field tends to zero as expected but slowly compared with the classical 
case. In fact, $E(r)$ behaves as $r^{-1}$ in clearly deviation with respect to the result shown in Eq.\ref{edr}.
Finally, the curvature and Kretschmann scalars hold the same asymptotic behaviour of the results obtained in absence of running,
i.e. $R\sim r^{-4}$ and $\mathcal{K}\sim r^{-6}$.

\subsection{Horizons}
Applying the condition $f(r_H)=0$ one obtains the scale dependent horizon which reads
\begin{align}
r_{H} &= - \frac{1}{\epsilon} \bigg[ 1- \Bigl[1 + 3 \epsilon r_0 \Bigl]^{1/3} \bigg],
\\
r_{\pm} &= -\frac{1}{\epsilon} \Bigg[1 + \frac{1}{2}(1 \pm i \sqrt{3})\bigg[1 + 3\epsilon r_0 \bigg]^{1/3}\Bigg].
\end{align}
where $r_0$ is the classical value given by Eq. \ref{rclasico34}. Note that one obtains three horizons, out of which one is
real (physical horizon)
and two $r_{\pm}$ are complex (non-physical).

\begin{figure}[ht!]
\centering
\includegraphics[width=\linewidth]{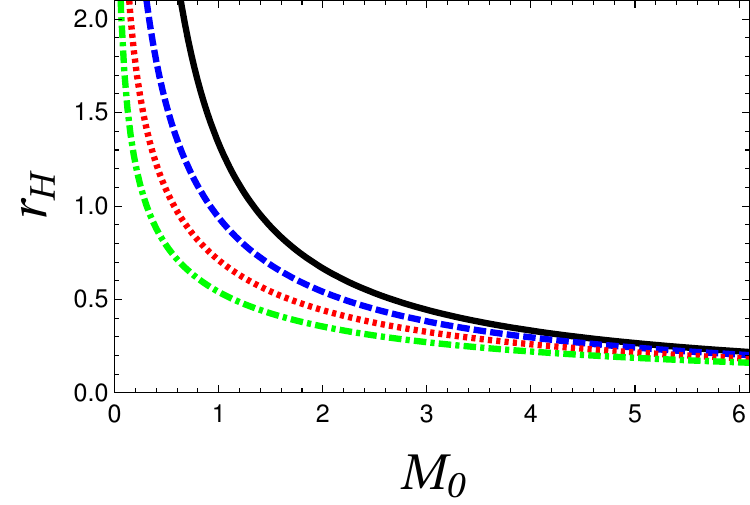}
\caption{\label{rH_I} 
Black hole horizons $r_H$ as a function of the mass $M_0$ for 
 $\epsilon=0$ (black solid line), $\epsilon=0.4$ (blue dashed line), $\epsilon=1$ (dotted red line) and $\epsilon=2$ (dotted dashed green line). 
The values of the rest of the parameters have been taken as unity.
}
\end{figure}

\noindent In addition, since the scale dependence of coupling constants is usually assumed to be weak, it is reasonable
to consider the dimensionful parameter $\epsilon$ as small compared to the other scales and, therefore, one can expand around $\epsilon$ close to zero, which 
gives us
\begin{align}
r_H &\cong  r_0 \bigg[ 1 - \epsilon r_0 + \frac{5}{3} \  (\epsilon r_0)^2 + \cdots \bigg].
\end{align}
One should note that when $\epsilon$ tends to zero the classical case is recovered. Besides, although $\epsilon$ could 
take positive or negative values, here in order to obtain desirable physical results we require that $\epsilon > 0$.
In our set of solutions $\{G(r), E(r), f(r), e(r)^3\}$ we can expand around zero for small values of $\epsilon$, i.e.
\begin{align}
G(r) &\approx G_0\Bigl[1 -r \epsilon + \mathcal{O}(\epsilon^2)\Bigl],\label{uno}
\\
E(r)&\approx E_0(r) + \mathcal{O}(\epsilon^2),\label{dos}
\\
f(r) &\approx f_0(r) + \bigg[2 G_0 M_0 - \frac{4 G_0 Q_0^2}{r} \bigg] r \epsilon  + \mathcal{O}(\epsilon^2), \label{tres}
\\
e^3(r) &\approx e_0^3 \bigg[1 + \mathcal{O}(\epsilon^2) \bigg].\label{cuatro}
\end{align}

\subsection{Thermodynamic properties}
Using the horizon structure and the lapse function (which is given by Eq. \ref{setsolu}) one can calculate the Hawking temperature of the corresponding scale dependent black hole. At the outer horizon this temperature is given by the simple formula
\begin{align}
T_H &= \frac{1}{4 \pi} \Bigg|\lim_{r\rightarrow r_H} \frac{\partial_r g_{tt}}{\sqrt{-g_{tt}g_{rr}}} \Bigg|,
\end{align}
which reads in term of the horizon radius
\begin{align} \label{TH}
T_{H} &= \frac{1}{4 \pi} \Bigg| \frac{M_0 G_0}{r_H (1+ \epsilon r_H)} 
\Bigg|.
\end{align}

\begin{figure}[ht!]
\centering
\includegraphics[width=\linewidth]{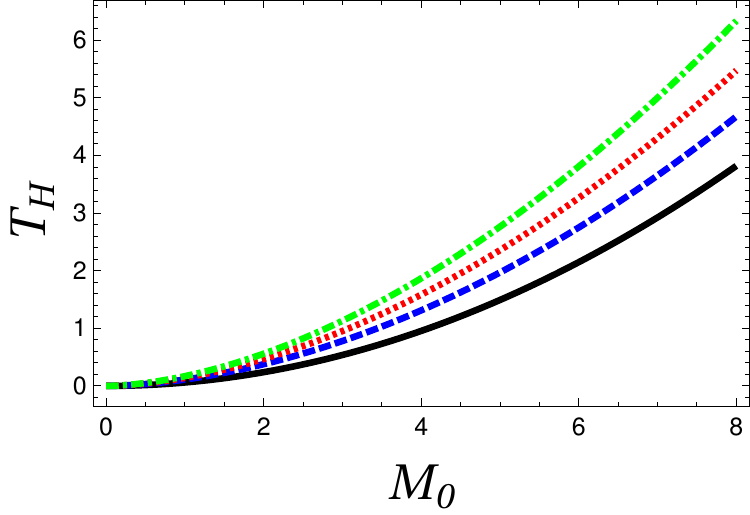}
\caption{\label{TH_I} 
Hawking temperature $T_H$ as a function of the classical mass $M_0$ for $\epsilon=0$ (black solid line), 
$\epsilon=20$ (blue dashed line), $\epsilon=50$ (dotted red line) and $\epsilon=100$ (dotted dashed green line). The values of 
the rest of the parameters have been taken as unity.
}
\end{figure}
\noindent
In order to recover the classical result we expand around $\epsilon = 0$ and upon evaluating at 
the classical horizon we obtain
\begin{align}
T_H(r_H) &\approx T_0(r_0) \Bigg| 1 + \frac{1}{3} (\epsilon r_0)^2 + \mathcal{O}(\epsilon^3)\Bigg|  ,
\end{align}
where it is clear that $\epsilon \rightarrow 0$ coincides with Eq. \ref{Tclassic} as it should be.
\\
In addition, the Bekenstein-Hawking entropy obeys the well-known relation heritage of Brans-Dickey theory
applied to the (2+1)-dimensional case
\begin{align}
S &= \frac{1}{4} \oint  {\mathrm {d}}x \frac{\sqrt{h}}{G(x)},
\end{align}
where $h_{ij}$ is the induced metric at the horizon. For the present circularly symmetric solution this integral is trivial 
because the induced metric 
for constant $t$ and $r$ slices is $\mathrm {d}s = r\mathrm {d}\phi$ and moreover $G(x) = G(r_H )$ is constant along the
horizon. Using these facts, the entropy for this solution is found to be \cite{angel,Rincon:2017ypd}
\begin{align}\label{S}
S &= \frac{\mathcal{A}_H}{4 G(r_H)} = S_0(r_{H}) (1 + \epsilon r_H),
\end{align}
while for small values of $\epsilon$ one obtains
\begin{align}
S &\approx S_0(r_0)
\Bigg[
1 -
\frac{1}{3}(\epsilon r_0)^2 
+ \mathcal{O}(\epsilon^3)
\Bigg],
\end{align}
which, of course, coincides with the classical results in the limit 
$\epsilon \rightarrow 0$.
\begin{figure}[ht!]
\centering
\includegraphics[width=\linewidth]{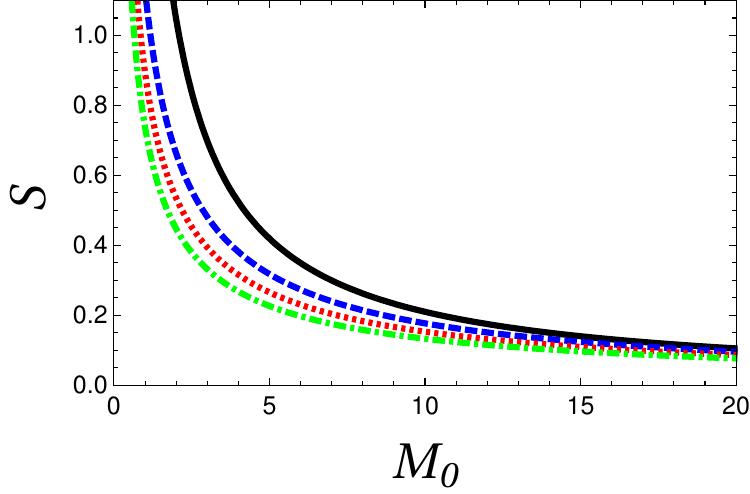}
\caption{\label{S_I} 
The Bekenstein-Hawking entropy $S$ as a function of the classical mass $M_0$ for $\epsilon=0$ (black solid line), $\epsilon=20$ (blue dashed line), $\epsilon=50$ (dotted red line) and $\epsilon=100$ (dotted dashed green line). The other values have been taken as unity.
}
\end{figure}
In addition, the heat capacity (at constant charge) $C_{Q}$ can be calculated by
\begin{align}
C_{Q} = T \ \frac{\partial S}{\partial T}\Bigg|_{Q}.
\end{align} 
Combining Eq. \ref{TH} with \ref{S} one obtains the simple relation
\begin{align}
C_Q &= -\frac{1}{8}\frac{M_0}{T_H} = -S_0(r_{H}) (1 + \epsilon r_H).
\end{align}
Note that the black hole is unstable since $C_Q < 0$, and it coincides with the classical result in the limit $\epsilon \rightarrow 0$.

\subsection{Total charge}
As in the previous case, the total charge $Q$ needs to be computed by the relation \cite{Carroll:2004st}
\begin{align}
Q &= \int \sqrt{-g} \mathrm{d} \Omega \Bigg(\frac{\mathcal{L}_F F_{\mu \nu}}{e_k^{2 \beta}}\Bigg)n^{\mu} \sigma^{\nu}.
\end{align}
In this case we obtain
\begin{align}
Q &=\frac{Q_{0}}{2 e_{0}^{3/2}},
\end{align}
which also is proportional to the classical value and does not have $\epsilon$ dependence.

\section{Discussion}\label{Discussion}

Scale dependent gravitational couplings can induce non-trivial deviations from classical Black Holes solutions.
We have studied two cases, first Einstein-Maxwell and second Einstein-power-Maxwell case. Both of them have a common feature: the lapse function tends to zero when $r \rightarrow \infty$, characteristic which is absent in the classical solutions.
In addition, the total charge is modified
as a consequence of our scale dependent framework. Moreover, we have found that, for the same value of the classical black hole mass, the apparent horizon radius (and the Bekenstein-Hawking entropy) decreases when the strength of the scale dependence increases. This is in agreement with the findings in  \cite{Bonanno:2000ep,Bonanno:2006eu,Reuter:2006rg,Reuter:2010xb,Falls:2012nd,Cai:2010zh,Becker:2012js,Becker:2012jx,Koch:2013owa,Koch:2013rwa,Ward:2006vw,Burschil:2009va,Falls:2010he,Koch:2014cqa,Bonanno:2016dyv}.
On the other hand, the Hawking temperature increases with $\epsilon$. Please, note that the effect of the scale dependence in the Einstein-power-Maxwell case is stronger than the Eintein-Maxwell case. 
The behaviour of the electromagnetic coupling $e(r)$ depends on the choice of the electromagnetic Lagrangian density. While $e(r)$ goes to zero in the limit $r \to \infty$ for a Maxwell Lagrangian density, it approaches a constant value for the power-Maxwell case.
Finally, it is well known that a black hole (as a thermodynamical system) is locally stable if its heat
capacity is positive \cite{Dehghani:2016agl}. In both scale dependent cases it is found that these black holes are unstable ($C_Q < 0$), like their classical counterparts. 

\section{Conclusion}\label{Conclusion}
In this article we have studied the scale dependence of charged
black holes in three-dimensional spacetime both in linear (Einstein-Maxwell) and 
non-linear (Eins\-tein-power-Maxwell) electrodynamics. In the second ca\-se we have considered the 
case where the electromagnetic energy momentum tensor is traceless, which happens for
$\beta=3/4$. After presenting the models
and the classical black hole solutions, we have allowed for a scale dependence of the 
electromagnetic as well as the gravitational coupling, and we have solved the corresponding
generalized field equations by imposing the "null energy condition" in three-dimensional
spacetimes with static circular symmetry. Horizon structure, asymptotic spacetimes
and thermodynamics have been discussed in detail. 

\section{Appendix}\label{Appendix}

In this appendix we study some features of the scale dependent $(2+1)$ gravity coupled to a power-Maxwell source
for an arbitraty $\beta$. For this system the action is given by
\begin{eqnarray}\label{acgen}
S= \int d^{3}x\sqrt{-g}\bigg[\frac{1}{16\pi G(r)}R - \frac{1}{e(r)^{2\beta}}\mathcal{L}(F)\bigg],
\end{eqnarray}
where $G(r)$ and $e(r)$ are the gravitational and the the electromagnetic scale-dependent couplings, $R$ is the Ri\-cci scalar, 
$\mathcal{L}(F )=C^{\beta}|F |^{\beta}$ is the electromagnetic Lagrangian density, 
$F=(1/4)F_{\mu\nu}F^{\mu\nu}$ is the Maxwell invariant, and $C$ is a dimensionless constant which depends on the
choice of $\beta$.  Metric signature $(-, +, +)$ and natural
units $(c =\hbar=k_{B}= 1)$ are used in our computations.\\ 
Variations of the Eq.\ref{acgen} with respect to the metric field lead to the modified Einstein's equations
\begin{eqnarray}\label{eingen}
R_{\mu\nu}-\frac{1}{2}g_{\mu\nu}R=8\pi\frac{G(r)}{e^{2\beta}(r)}T_{\mu\nu}-\Delta t_{\mu\nu}, 
\end{eqnarray}
where $T_{\mu\nu}$ stands for the power-Maxwell energy momentum tensor and
\begin{eqnarray}
\Delta t_{\mu\nu}=G(r)\Bigl(g_{\mu\nu}\square -\nabla_{\mu}\nabla_{\nu}\Bigl)G(r)^{-1},
\end{eqnarray}
is the non-material energy momentum tensor which ari\-ses as a consecuence of the scale dependence of the gra\-vitational coupling. 
On the other hand, after variations of the action Eq.\ref{acgen} with respect to the electromagnetic four--potential, $A_{\mu}$, one obtains 
the modified Maxwell equations
\begin{eqnarray}\label{ecpM}
D_{\mu}\Bigg(\frac{\mathcal{L}_{F}F^{\mu\nu}}{e(r)^{2\beta}}\Bigg)=0. 
\end{eqnarray}
Henceforth, only static and circularly symmetric solutions will be considered. Therefore we shall assume the ansatz 
\begin{align}
ds^{2}&=-f(r)dt^{2}+f(r)^{-1}dr^{2}+r^{2}d\Omega^{2}, \\ 
F_{\mu\nu}&=(\delta^{t}_{\mu}\delta^{r}_{\nu}-\delta^{r}_{\mu}\delta^{t}_{\nu})E(r),
\end{align}
for the metric and the electromagnetic tensor, respectively. With the former prescription is straightforward to prove, from Eq.\ref{ecpM}, that 
the electric field is given in terms of the electromagnetic coupling by
\begin{eqnarray}\label{genE}
E(r)= \frac{2^{\frac{\beta -1}{2 \beta -1}} C^{-\frac{\beta }{2 \beta -1}} Q_{0}^{\frac{1}{2 \beta -1}} e(r)^{\frac{2 \beta }{2 \beta -1}}}{\beta ^{\frac{1}{2 \beta -1}}
r^{\frac{1}{2 \beta -1}}},
\end{eqnarray}
or, in a more convenient way
\begin{eqnarray}\label{genE2}
E(r)= \Bigg[ 
\Bigg(\frac{2^{\beta-1} C^{-\beta}}{\beta}\Bigg)
 \Bigg(\frac{Q_0}{r}e(r)^{2\beta}\Bigg)
\Bigg] ^
 {\frac{1}{2\beta -1}}.
\end{eqnarray}
Please, note that setting $\beta=1$ and $C=1$ the electric field reported in Eq.\ref{setsoluII} is recovered 
\begin{eqnarray}
E(r)=\frac{Q_{0}}{r}e(r)^2 .
\end{eqnarray}
In the same way, for $\beta=3/4$ and $C^{3/4}=2^{7/3} 3^{-\frac{4}{3}}e_{0}^2 Q_0^{2/3}$ one obtain
\begin{eqnarray}
E(r)=\frac{Q_{0}}{r^2}\bigg(\frac{e(r)}{e_{0}}\bigg)^{3},
\end{eqnarray}
in complete agreement with Eq.\ref{setsolu}. It is worth noting that, even in the general case the electric field depends on an specific power of
the charge as a consequence of the non--linear electrodynamics, in the cases $\beta=1$ and $\beta=3/4$, this behaviour is not observed due to a 
particular setting of $C$. \\ 
If the null energy condition is used as an additional condition, we obtain that the scale-dependent gravitational coupling reads
\begin{eqnarray}\label{genG}
G(r)=\frac{G_{0}}{1+\epsilon r},
\end{eqnarray}
where $G_{0}$ is Newton's constant and $\epsilon$ is the running parameter. Note that the classical limit is recovered in the limit $\epsilon\to 0$.
Finally, Eq.\ref{eingen} reduces to a pair of differential equations for
$\{f(r),e(r)^{2\alpha}\}$ given by
\begin{align}
&2^{\alpha} \kappa_0 C^{-\alpha } Q_0^{2 \alpha } (2 \beta -1) r e(r)^{2 \alpha } \  +\nonumber\\
&\beta ^{2 \alpha } r^{2 \alpha } \Bigl((2 r \epsilon +1) f'(r)+2 \epsilon  f(r)\Bigl)=0,\label{una}\\
&2^{\alpha} \kappa_0 C^{-\alpha } Q_0^{2 \alpha } e(r)^{2 \alpha } \ - \nonumber\\
&\beta ^{2 \alpha } r^{2 \alpha } \Bigl((r \epsilon +1) f''(r)+2 \epsilon  f'(r)\Bigl)=0,\label{dos}
\end{align}
where $\alpha=\frac{\beta}{2\beta-1}$ and $\kappa_0 = 8 \pi G_0$. 
It can be checked by the reader that, in the case $\beta=3/4$, the solutions of the set of equations \ref{una}, \ref{dos}, \ref{genG} and \ref{genE}
coincide with those listed in Eq.\ref{setsolu} after an appropriate choice of the integration constants.

\begin{acknowledgements}
The author A.R. was supported by the CONICYT-PCHA/\- Doctorado Nacional/2015-21151658.
The author P.B. was supported by the Faculty of Science and Vicerrector\'{\i}a de Investigaciones of Universidad de los
Andes, Bogot\'a, Colombia.\\
The author B.K. was supported by the Fondecyt 1161150.\\
The author G.P. acknowledges the support from "Funda{\c c}{\~a}o para a Ci{\^e}ncia e Tecnologia". 
\end{acknowledgements}


\bibliographystyle{unsrt}         

\end{document}